\documentclass[prl,twocolumn,floatfix,epsf,psfig]{revtex4}
\pdfoutput=1 %% this is very important command to upload pdf figures in condmat
\usepackage{graphicx}
\renewcommand{\figurename}{Figure}
\usepackage{amsmath}

\begin{document}
%\title{AuSn: a 2D surface alloy grown on Au(111)} 
%\title{Linear bands at zone center in a non-honeycomb lattice} 
%\title{Dirac cone in a non-honeycomb lattice} 
\title{Dirac cone in a non-honeycomb surface alloy} 
\author{Pampa Sadhukhan$^1$, Dhanshree Pandey$^{2,3}$, Vipin  Kumar Singh$^1$, Shuvam Sarkar$^1$, \\ Abhishek Rai$^1$, Kuntala Bhattacharya$^4$,  Aparna Chakrabarti$^{2,3}$ and Sudipta Roy Barman$^{1}$}
\affiliation{$^1$UGC-DAE Consortium for Scientific Research, Khandwa Road, Indore 452001, Madhya Pradesh, India}
\affiliation{$^2$Homi Bhabha National Institute, Training School Complex, Anushakti Nagar, Mumbai  400094, Maharashtra, India}
\affiliation{$^3$Theory and Simulations Laboratory,  Raja Ramanna Centre for Advanced Technology, Indore 452013, Madhya Pradesh, India}
\affiliation{$^4$Department of Physics, Indian Institute of Space Science and Technology, Thiruvananthapuram 695547, Kerala, India}

\begin{abstract}
	We demonstrate unexpected occurrence of linear bands resembling  Dirac cone at the zone-center   of Au$_2$Sn surface alloy with  $\left( \begin{smallmatrix} 2&1\\ 1&3 \end{smallmatrix} \right)$ surface structure formed by deposition of about 0.9 ML Sn on Au(111) at elevated temperature. The surface exhibits an oblique symmetry with unequal lattice constants making it the first two dimensional surface alloy to exhibit Dirac cone with a non-honeycomb lattice. 
\end{abstract}

%\pacs{~79.60.-i} %%Photoemission spectroscopy of solids and liquids 
\maketitle 
% For (1x1) B.Z: M=1.258 A, K=1.4518 A For (2113) B.Z: M'=0.436 A, K'=0.7057 A with best regards, Pampa

%**ask lalla about tem mesh for free standing ** 
%\section{Introduction}
          Since the discovery of graphene\cite{Novoselov05,Geim07}, fabrication of atomically thin two dimensional (2D)  materials with nontrivial band topology  has attracted enormous attention primarily because of their dissipationless conduction.  This led to the discovery of a family of  2D quantum materials  with exotic properties; for example, to mention a few are stanene\cite{Xu13}, silicene\cite{Vogt12}, aluminene\cite{Kamal15}, borophene\cite{Mannix15} and phosphorene\cite{L.Li14}.  In particular, the prediction of stanene\cite{Xu13}, a buckled tin honeycomb layer, to be a quantum spin Hall insulator with Dirac cone-like linear energy dispersion and a large gap of 0.3 eV  has stirred up  efforts to realize it on substrates.  The first experimental synthesis of stanene was obtained by molecular beam epitaxy  on Bi$_2$Te$_3$ substrate\cite{Zhu15}. Recently, stanene with a even larger gap of  0.44 eV   was reported on  InSb(111)\cite{Xu17}. Two very recent studies\cite{Yuhara18,Deng18} demonstrated existence of stanene on metal substrates.  Deng  $et~al.$\cite{Deng18} reported epitaxial growth of flat stanene on Cu(111)  and obtained $s-p$ band inversion as well as spin-orbit coupling (soc) induced topological gap. Yuhara $et~al.$\cite{Yuhara18} also reported planar  stanene  on  Ag$_2$Sn surface alloy on  Ag(111), however angle resolved photoemission (ARPES) showed a parabolic band dispersion. % due to hybridization with the underlying surface alloy. 

        Gold is an interesting substrate for stanene growth. It also exhibits a Rashba spin-orbit split surface state in the $L$-gap\cite{Lashell96, Reinert01}. 
        ~However, the existing literature of Sn growth on Au(111) presents conflicting results. A density functional theory (DFT) work predicted that a planar stanene is  energetically favorable\cite{Nigam15}; while another study showed that its band structure  would be modified due to bonding with Au substrate\cite{Guo16}. In contrast, there are both experimental\cite{Barthes81,Zhang91} as well as theoretical studies\cite{Mier17,Canzian10}  that show occurrence of Au-Sn surface alloy at room temperature (RT). %While surface alloying however would not favor honeycomb stanene structure, it could also be interesting because surface alloys such as  Bi-Ag\cite{Ast07}, Pb-Au\cite{Chen15} show very large Rashba splitting. 
        ~We have studied the growth of Sn on Au(111) under different conditions and demonstrate  presence of perfectly linear Dirac-like bands crossing the Fermi level ($E_F$) at the zone center ($\overline\Gamma$) with  Fermi velocity comparable to graphene. The Dirac-like bands occur only in a specific Au-Sn $\left( \begin{smallmatrix} 2&1\\ 1&3 \end{smallmatrix} \right)$  surface alloy phase with oblique symmetry  ($\gamma$ =70.9$^\circ$ and $b$:$a$= $\sqrt{7}/\sqrt{3}$) that is formed at high temperature, but is stable when cooled to RT and has a composition of Au$_2$Sn. Our DFT calculation for a model structure, namely modified Lieb lattice with oblique symmetry shows presence of  linear bands for this binary alloy.
      
%\section{Experimental details}
 Polished and oriented Au(111) crystal %from Mateck GmbH, Germany 
 ~was cleaned in-situ by repeated cycles of 0.5 keV Ar$^{+}$ ion sputtering for 15 min followed by annealing at 673~K for about 10 min. %Sharp 1$\times$1 LEED pattern was monitored to check well ordered and clean substrate surface. 
 ~Sn was deposited % at the rate of 0.05 ML/min 
 ~using a water cooled Knudsen cell\cite{Shukla04} operated %at the base pressure of 8$\times$10$^{-10}$mbar and 
 ~at 1078~K% 1003~K(we can write ARPES Sn cell slow rate deposition temperature i.e 1003K and 1078K is STM Sn cell temperature)..pampa. %During deposition substrate temperature (T$_{S}$) was kept fixed at the temperature varying from RT to 523K. 
 %deposition is referred in terms of monolayer(ML) based on direct observation from STM  images by measuring the fractional area of the substrate covered by atomic hight island. The STM image analysis has been performed by using SPIP 6.4.1 software.
 ~Low energy electron diffraction (LEED) was performed using a four grid rear view optics.% from OCI Vacuum Microengineering. 
 ~The STM measurements were carried out in a variable temperature STM  work station %from Omicron Nano Technology. The STM images were recorded 
 ~in the constant current mode by %keeping the specimen at ground potential and 
 ~applying the bias to a tungsten tip. The coverage has been determined from the change in slope of the Auger electron spectroscopy signals from  Sn and Au as well as by STM. %The base pressure of the chamber was 2$\times$10$^{-11}$ mbar.
  %Cleaniness of the bare surface was checked by X-ray photoelectron spectroscopy using R4000 electron energy analyser and monochromatized AlK$\alpha$ laboratory x-ray source with 1486.6 eV photon energy from Scienta GmbH at the base pressure of 1$\times$10$^{-10}$ mbar. 
     % X-ray photoelectron spectroscopy (XPS)  was carried out in a workstation supplied by Prevac, Poland using R4000 electron energy analyser and monochromatized AlK$\alpha$ laboratory x-ray source with 1486.6 eV photon energy from Scienta GmbH at the base pressure of 1$\times$10$^{-10}$ mbar .  The spectra were recorded at normal emission geometry with 100 eV pass energy, transmission lens mode and 0.3 mm slit width. The instrumental resolution was 0.335 eV at room temperature, obtained  by fitting the Au fermi edge with a Gaussian function. 
~            The photoemission measurements were carried out in a separate workstation %supplied by Prevac 
~ using R4000 electron energy analyzer.% from Scienta GmbH. 
~The base pressure of both the workstations were better than 2$\times$10$^{-10}$ mbar.
      % Monochromatized Al-K$\alpha$ x-ray and UPS  source of photon energy 1486.6 eV, 21.2 and 40.8 eV were used. In ARPES, 
~      For ARPES,  the overall energy resolution measured by fitting the Au Fermi level including RT  broadening was 100~meV, while the angular resolution was 1\,$^{\circ}$ for  acceptance angle of  $\pm$15$^\circ$.  The energy resolution for XPS using  monochromatized Al K$\alpha$ source was  0.34~eV. %x-ray All the photoemission measurements are performed at RT, normal emission geometry using  Angular14, angular30 lens mode with acceptance angle $\pm$ 7.5$^\circ$ and $\pm$ 15$^\circ$ 10 eV pass energy for ARPES and Transmission lens mode,100 eV pass energy for XPS.  
      ~ The core-level spectra have been fitted using a least square error minimization procedure  where % where as Sn 4d clearly shows a broad spectra(fig.3(b,d)) which indicates more than one components is present for both the cases. 
      ~Doniach-$\breve{S}$unji$\acute{c}$ (DS) line shape\cite{Doniach70} convoluted with a Gaussian function representing the instrumental broadening has been used to represent each component. % The width of each component was kept fixed at the bulk metal values.  

     The electronic structure calculations using density functional theory (DFT) have been performed by Vienna $ab~initio$ simulation package (VASP)\cite{VASP} using  the projector  augmented wave  method\cite{PAW}.    For exchange-correlation functional, the generalized gradient approximation  %given by Perdew, Burke, and Ernzerhof (PBE) 
     ~has been employed\cite{PBE}. We use an energy cutoff of 350 eV for the plane waves. The final energies have been calculated with a $k$ mesh of 29$\times$21$\times$1. %We simulate the $2D$  Lieb lattice with two Sn and four Au atoms, respectively, arranged in a specific  order with typically 
~A vacuum region of about 18 \AA~ is considered in the $z$-direction. %(direction perpendicular to the plane of the Lieb lattice) 
%~to avoid the interaction between two adjacent unit cells in the periodic arrangement.
~The energy and the force tolerance for our  calculations are 1 $\mu$eV and 20 meV/\AA, respectively.

 \begin{figure}[t]
 		\vskip -5mm
		\includegraphics[width=85mm]{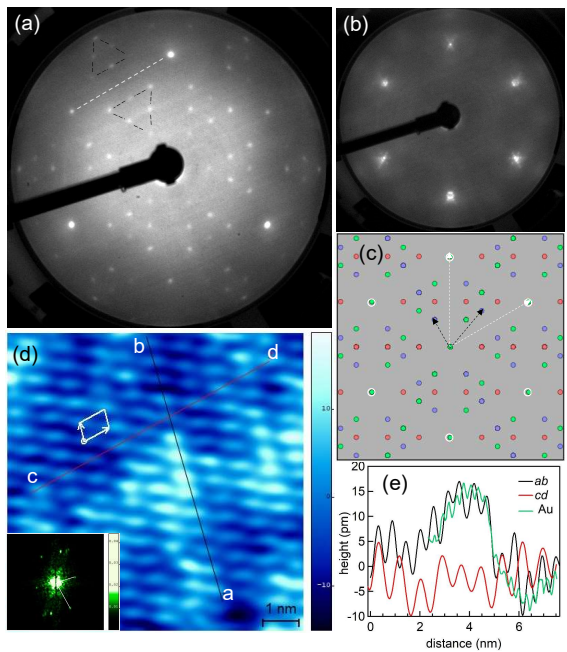} %SOURCE: PROJECTS\Sn_Au_111\analysis\PS\PS_Sn_Au_XPS_fit_ver1_srb.pxp, Layout 5	
 	\vskip -20mm
 	 	\caption{The low energy electron diffraction (LEED) pattern with 104 eV beam energy of (a) the (2113) phase of  Sn/Au(111) obtained by depositing $\approx$0.9 ML Sn at 413~K  and subsequently cooled to RT and (b) the bare  Au(111) substrate.   (c) A simulation of the  LEED pattern in (a) using the LEEDpat\cite{LEEDpat}, the three domains rotated by 120$^{\circ}$ are shown by green, blue and red circles, while the white circles represent their common spots that appear at (1$\times$1) position. The primitive reciprocal lattice vectors for the blue domain are marked by black dashes, while the white dashes represent those of Au(111). (d) A high resolution %atomic scale 
 	 		~scanning tunneling microscopy (STM) image showing one domain of the (2113) phase using tunneling current of 0.9~nA and a bias voltage of -0.3~V. The   real space primitive  unit cell and basis vectors (white arrows) are shown, inset shows the Fourier transform of this image. (e) Height profiles along ab (black line) and cd (red line) of (d) are compared with the height profile (green line) along an atomic array perpendicular to the discommensurate lines of Au(111) herringbone reconstruction.} 
 	\label{leedstm}
 \end{figure}
    
%   \section{RESULTS AND DISCUSSIONS} 
  % \subsection{Atomic Structure}
   
\begin{figure}[t]
	%\epsfxsize=50mm
	%\centering
	\includegraphics[width=85mm]{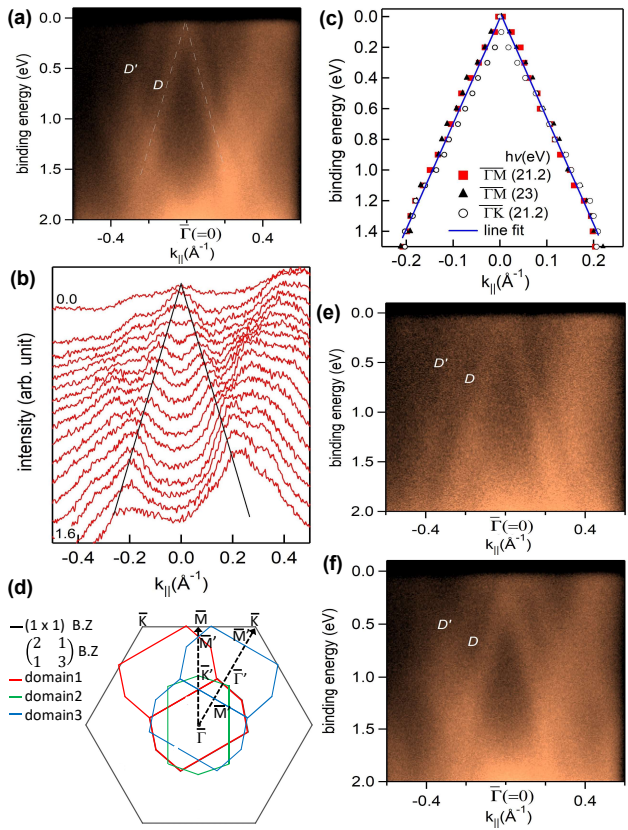} %SOURCE: PROJECTS\Sn_Au_111\analysis\PS\PS_Sn_Au_XPS_fit_ver1_srb.pxp, Layout 5	
	\caption{ (a) ARPES spectra of the (2113) Sn/Au(111) phase using 21.2~eV photon energy ($h\nu$)   along the $\overline{\Gamma}$-$\overline{M}$ direction. The linear band $D$ is highlighted by white dashed lines.  (b) The momentum distribution (MDC) curves from $E_B$= 0 to 1.6 eV at step of 0.1 eV in (a).  (c) $E_B$  as a function of $k_\parallel$ (red squares) for $D$ band obtained from the MDC curves in (b). (d) The first Brillouin zones of Au(111) (outer regular hexagon) and the (2113) phase (inner elongated hexagon) with the high symmetry points indicated.  ARPES spectra along (e) $\overline{\Gamma}$-$\overline{M}$  with $h\nu$= 23 eV and (f)  along $\overline{\Gamma}$-$\overline{K}$ with $h\nu$= 21.2 eV.} 
	\label{arpes}
\end{figure}

The LEED pattern of $\approx$0.9~ML Sn/Au(111) deposited at a substrate temperature ($T_S$) of 413~K in Fig.~\ref{leedstm}(a) is completely different from the Au(111) substrate in Fig.~\ref{leedstm}(b). Au(111) exhibits satellite spots around each 1$\times$1 spot that are  related to the 22$\times$$\sqrt{3}$  reconstruction. On the other hand, the Sn/Au(111) pattern comprises of characteristic pairs of triangles (black dashed) each formed by six spots. These  have reflection symmetry across the line (white dashed) joining  two adjacent 1$\times$1 spots. The absence of the satellite spots related to the bare substrate reconstruction obviously means formation of a totally different surface structure. We have simulated the LEED pattern considering a matrix\cite{LEEDpat}
$M = \left( \begin{smallmatrix} 2&1\\ 1&3 \end{smallmatrix} \right)$,
%matrix with square brackets
%$\bf M= \begin{bmatrix}
%2 & 1\\
%1 & 3
%\end{bmatrix}$
~the three sets of  red, green and blue colored spots correspond to three 120$^{\circ}$ rotated domains (Fig.~\ref{leedstm}(c)). We henceforth denote this structure by (2113).    The reciprocal unit cell vectors  for one of the domains  and the substrate are shown in Fig.~\ref{leedstm}(c). If {\bf $a_1$} (={\bf $a_2$}) is the substrate direct lattice vector, the overlayer direct lattice vectors shown by white arrows in Fig.~\ref{arpes}(d) are {\bf $b_1$}= $\sqrt{3}$$\times$$a_1$, while {\bf $b_2$}= $\sqrt{7}$$\times$$a_2$. {\bf $b_1$} ({\bf $b_2$}) is rotated by 30$^{\circ}$ (100.9$^{\circ}$) with respect to {\bf $a_1$} and thus the unit cell of the overlayer is oblique with %the direct lattice vectors $b_1$= 4.99 \AA, $b_2$= 7.62\,\AA~ and 
~ $\gamma$= 70.9$^{\circ}$. %srb The (2113) phase described here can be obtained by  deposition of $\approx$0.9 ML Sn in a  $T_S$ range of 393-413~K. An alternative way to obtain this phase is to deposit similar amount of Sn on Au(111) at  RT and subsequently anneal to $T_S$= 393-413~K for 30 min.

   A high resolution atomic scale STM image in Fig.~\ref{leedstm}(d) shows the oblique mesh, where the lengths of the unit cell vectors {\bf $b_2$} and {\bf $b_1$} are estimated to be  8.4$\pm$0.3\,\AA~ and 5.5$\pm$0.3\,\AA, respectively, which are in the ratio of $\sqrt{7}/\sqrt{3}$. 
~The oblique symmetry of the unit cell is also evident with $\gamma$ having similar value as obtained from LEED (70$\pm$3$^{\circ}$). %as well as the lattice parameters are in fair agreement with the values obtained from  LEED.  %:  $a$= 4.99 \AA, $b$= 7.62\,\AA~ and  $\gamma$= 70.89$^{\circ}$,  considering the lattice constant of Au triangular mesh to be 2.88\AA. some difference could be ascribed to presence of thermal drift in STM. %, while the LEED values could be larger if there is any lattice relaxation of the Au substrate due to alloying 
~ The height profiles in Fig.\ref{leedstm}(e) %along the dashed white edges of the centered elongated hexagon in Fig.\ref{leedstm} 
~show that the overlayer is flat with atomic corrugation of  $\pm$5~pm that is similar to the bare substrate. In contrast to the height profile along cd, the height profile along ab that is perpendicular to cd shows a broad undulation. This is very similar to that observed across the  discommensurate lines of the Au(111) herringbone reconstruction (green line in Fig.~\ref{leedstm}(e)) indicating that although there is no buckling as in stanene, a long range undulation of the Au(111) surface is present. An important conclusion from our LEED and STM studies is that Sn deposition on Au(111) at 413~K does not result in a %highly sought after 
~honeycomb structure.
%Thus,  the expected  buckling of stanene\cite{Xu,Zhu15} %of 0.86\AA  %is not observed.

   %\section{Angle resolved photoemission spectroscopy} 

In spite of not being stanene with honeycomb structure, the ARPES of the (2113) phase demonstrates an unexpected result: two branches of highly linear bands ($D$) forming a '$\Lambda$' shape, as shown by white dashed lines in Fig.~\ref{arpes}(a)). The bands are linear along $\overline{\Gamma}$-$\overline{M}$  over a binding energy ($E_B$)  range starting from  1.5 eV at $k_\parallel$= 0.2\AA$^{-1}$   to the Fermi level ($E_F$), where the two branches of '$\Lambda$' meet at %($E_B$, $k_\parallel$)= (0, 0)  $i.e.$ 
~the $\overline{\Gamma}$ point. Its linearity  is established from the momentum distribution curves (MDC) (Fig.\ref{arpes}(b)) from which $E_B$ as a function of $k_{\parallel}$(red squares) is obtained (Fig.\ref{arpes}(c)).  

The first Brillouin zone of the (2113) phase obtained from the unit cell determined above %(other domains are 120$^{\circ}$ rotated) 
~is overlaid on that of the substrate in Fig.~\ref{arpes}(d). ARPES measured up to the $\overline{M}$ point of the substrate BZ spans the  $\overline{K^{\prime}}$ and $\overline{M^{\prime}}$ points of the overlayer BZ when all the domains are considered, but no other linear band is observed at either $\overline{K^{\prime}}$ or $\overline{M^{\prime}}$ (see Fig. S1(a,b) of  Supplementary material (SM)\cite{supplement}). Similarly, ARPES spectra up to the $\overline{K}$ point of substrate BZ  spans the  $\overline{M^{\prime}}$ and $\overline{\Gamma^{\prime}}$ points of overlayer BZ, but no other linear bands are observed  (see Fig. S1(c,d)\cite{supplement}). The linear band is  observed only at the common zone center $\overline{\Gamma}$ of the three domains. The linear bands are not observed at $\overline{\Gamma^{\prime}}$ point of domain3 possibly because of overlap with bands related to the other two domains for which this is an arbitrary $k$ point and the proximity of intense $s-p$ band of Au. 

 The $D$ band is also observed at a different photon energy of 23~eV, the $E_B$($k_{\parallel}$) variation from MDC (solid black triangles in Fig.~\ref{arpes}(c)) is very similar to 21.2~eV photon energy showing that it is surface related. Moreover, as expected for a Dirac cone,  $D$ is unaffected by the variation of the azimuthal angle $e.g.$ from  the $\overline{\Gamma}$-$\overline{M}$ direction (Fig.~\ref{arpes}(a)) to $\overline{\Gamma}$-$\overline{K}$ direction  (Fig.~\ref{arpes}(f) and black open circles in Fig.~\ref{arpes}(c)). The ARPES data for the  intermediate  azimuthal angles are shown in Fig.~S2 of  SM\cite{supplement}.

A least square fitting %after averaging over  the peak positions 
~obtained from MDC curves of Fig.~\ref{arpes}(a,e,f) with a blue straight line  provides an excellent fit. The magnitude of the slopes ($dE_B$/$dk_{\parallel}$) of the left and the right branches are essentially same, 6.88$\pm$0.1 and 6.92$\pm$0.1, respectively. %Considering the average slope (6.92$\pm$0.11) and 
~Applying the relation $v_F$ = $\frac{1}{\hbar}\frac{dE_B}{dk_{\parallel}}$, the Fermi velocity ($v_F$)  turns out to be 1.05$\times$10$^6$ m/s, which is very similar  to that of graphene (1$\times$10$^6$ m/s)\cite{Zhang05}. %and is larger than  silicene (6$\times$10$^5$ m/s)\cite{Liu11}, Sn/InSb(001) (7.3$\times$10$^5$ m/s)\cite{Ohtsubo13} or other typical topological insulators $e.g.$ Bi$_{2}$Se$_{3}$ (5$\times$10$^5$ m/s)\cite{Zhang09}. %Although the Fermi velocity here similar to graphene showing  its  massless relativistic nature, 
~However, the important differences with graphene  is that the structure is non-honeycomb and a single Dirac-like cone is observed at the zone center.% $\overline{\Gamma}$. % and    Dirac cones have been  observed in 2D materials so far mostly in honeycomb lattice. 
~Theoretical studies have predicted their existence of such single Dirac cones non-honeycomb  structures. For example, the Lieb lattice with a square symmetry  hosts a topologically nontrivial $Z_2$ invariant insulating phase with a single Dirac cone per BZ\cite{Weeks10,Slot17}. A  MoS$_2$ allotrope having square-octagonal ring structure has been shown theoretically to exhibit a single Dirac cone at $E_F$ at the zone center and its $v_F$ is comparable to that of graphene\cite{Li14}.    Graphaynes that are rectangular show two  nonequivalent distorted Dirac cones\cite{Malko12}. We have used a modified Lieb lattice model for our DFT calculations, as discussed latter.

It is interesting to note that another weak linear band $D^{\prime}$ in Fig.~\ref{arpes}(a,e,f) is observed that is parallel to $D$ with a  $k_{\parallel}$ momentum offset of about 0.2\,\AA$^{-1}$. Both the branches of $D^{\prime}$ cross $E_F$ at 0.2\,\AA$^{-1}$. The  energy offset  between $D$  and $D^{\prime}$  is about 0.5~eV. It has been shown in an earlier study that Au intercalation in graphene  induces large Rasbha spin orbit coupling (soc) of 0.1 eV\cite{Marchenko12}.  Giant Rashba effect has also been observed in a AuPb binary surface alloy at the surface zone center with a Rashba parameter value of 4.45 eV\,\AA\cite{Chen15}. If we interpret $D^{\prime}$ to be due to Rashba soc, the Rashba parameter value here is 5 eV\AA, which is similar to AuPb. As in case of $D$, the $D^{\prime}$ band is also unaffected by change of photon energy (Fig.~\ref{arpes}(e)) and the azimuthal angle% of radial direction of measurement in the $k_x$-$k_y$ plane 
~(Fig.~\ref{arpes}(f) and Fig. S2\cite{supplement}).
%  In addition to this, a parabolic band with the minimum at the  $\Gamma$ point at $E_B$= 2.3 eV  crosses the Fermi level at 0.5\,$\AA{}$$^{-1}$. 

  \begin{figure}[t]
  	%\epsfxsize=50mm
  	%\centering
  	\includegraphics[width=85mm]{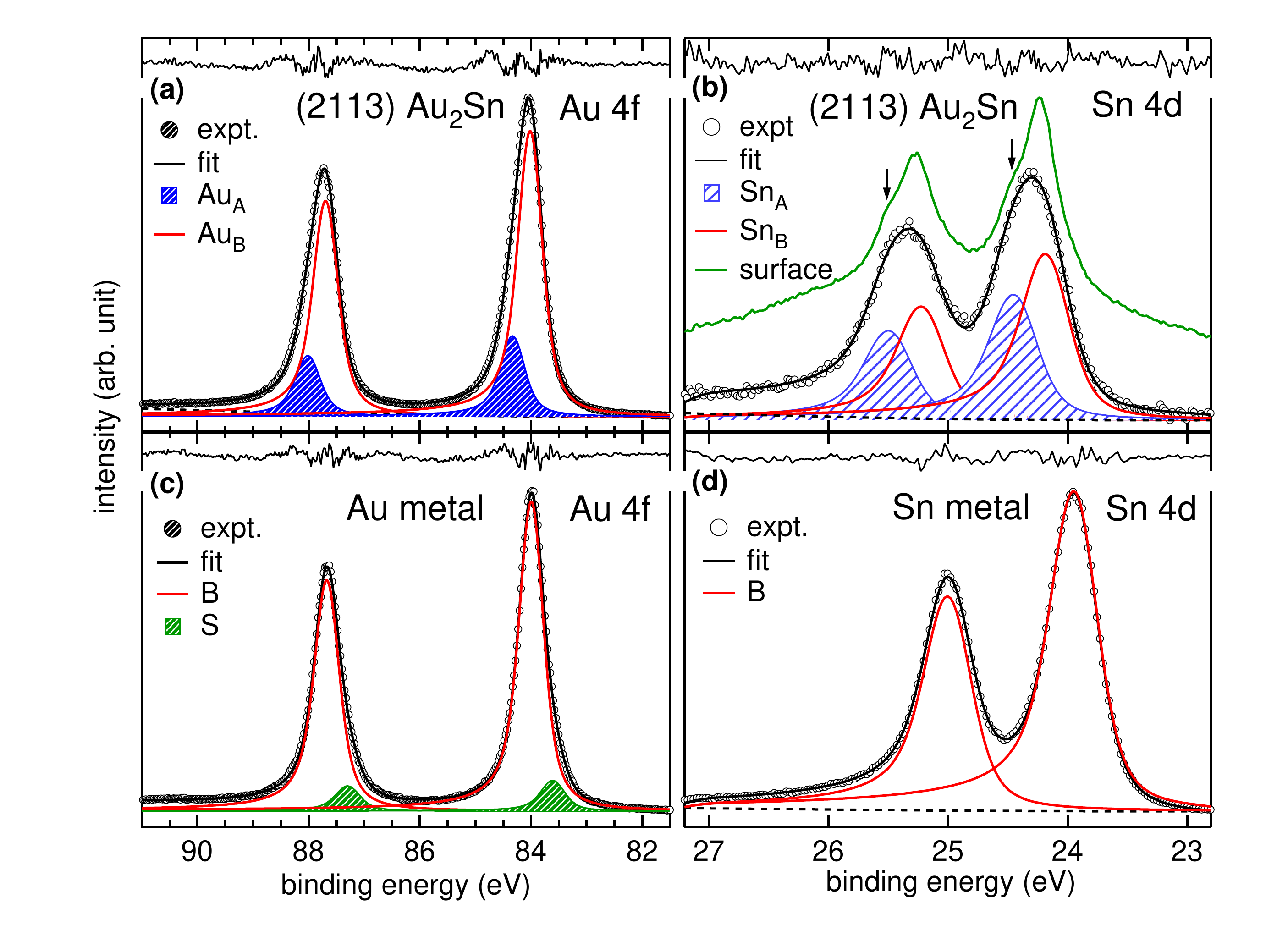} %SOURCE: PROJECTS\Sn_Au_111\analysis\PS\PS_Sn_Au_XPS_fit_ver1_srb.pxp, Layout 5	
  	\caption{ (a) Au 4$f$ and (b) Sn 4$d$  core level spectra of the (2113) phase compared with (c) Au 4$f$ (d) Sn 4$d$ of bulk metals Au and Sn, respectively. The residual of the least square fitting  (black line) is shown in the top of each panel. }
  	\label{core}
  \end{figure}

   %\subsection{XPS core-level spectra for 1ML Sn deposition at different substrate temperature}
  % \subsection{Electronic Structure}
   
Before proceeding further to understand the origin of the Dirac-like cones discussed above, we have  addressed the question of  alloying in the (2113) phase. % that forms at high temperature (423 K). 
%~in particular becasue there is no study in literature about high temperature Sn deposition on Au.%*here we might talk about our DAE paper* 
~Previous studies\cite{Barthes81,Zhang91} using LEED and AES  show formation of AuSn surface alloy at  RT. %, and thus  at $T_S$= 413~K surface alloying could be expected.  
~Au 4$f$ and Sn 4$d$ core-level spectra of the (2113) phase (Fig.~\ref{core}(a,b)) are compared with those of the corresponding  bulk metals  in Fig.~\ref{core}(c,d). 
~In Fig.~\ref{core}(c), Au 4$f$ shows the 4$f_{7/2}$ bulk peak at 84 eV, while the  surface component (shaded green) is shifted to lower $E_B$ at 83.6 eV\cite{Citrin78}. %The life time broadening  of the bulk peak is 0.3 eV, while the DS asymmetry parameter ($\alpha$)\cite{Doniach70} is 0.04. 
~In contrast, for the (2113) phase, the surface component is absent and both the Au 4$f$ spin-orbit split peaks exhibit an  asymmetry on the higher $E_B$  side that cannot be accounted for by DS asymmetry.  %because, if allowed to vary freely, $\alpha$ becomes  unphysically large and the fitting is poor. 
~A good quality fit is obtained only when an additional component ($\rm{Au}_A$, blue shaded) is considered, and its position is varied freely. The main peak ($\rm{Au}_B$) appears at 84~eV, while  $\rm{Au}_A$  appears at 0.3~eV higher $E_B$.    $\rm{Au}_B$ position coincides with the bulk component of Au metal (Fig.~\ref{core}(a,c)) and thus it can be assigned to the underlying substrate. On the other hand,  $\rm{Au}_A$ is related to the  (2113) 2D surface alloy. This is supported by an earlier work,  which show that the Au 4$f$ peak shifts to higher $E_B$ in bulk Au-Sn alloys compared to Au metal\cite{Friedman73}.   %$\rm{Ag}_A$ is related to the Au-Sn surface alloy. 

The Sn 4$d$ spectrum in Fig.~\ref{core}(d) for Sn metal  %simulated by a single component 
~shows  %, the life-time broadening and $\alpha$ are  0.2 eV and 0.09, respectively.  The 
~the 4$d_{5/2}$ and 4$d_{3/2}$  peaks  at 24~eV and 25~eV, respectively. In contrast,  the corresponding peaks in the (2113) phase are both shifted to higher $E_B$ by 0.3~eV. This clearly  indicates surface alloying since in bulk Au-Sn alloys, such shift  to higher $E_B$ is reported with respect to Sn metal\cite{Friedman73}. It is noted that the Sn 4$d$ peaks are broader and fitting with single component fails, indicating presence of at least two components.% since fitting with a single component does not give good fit and furthermore the life-time broadening is too large. 
~This is established by the Sn 4$d$ spectrum  recorded with higher resolution, % using UPS,  
~where the shoulder  depicting the second component is showed by an arrow (Fig.~\ref{core}(b), green line). Thus,  presence of two non-equivalent Sn atom positions in the Au-Sn alloy is indicated. These components ($\rm{Sn}_A$ and $\rm{Sn}_B$) are separated by 0.3 eV.  The composition of the surface alloy is determined to be Au$_2$Sn, considering the areas of $\rm{Au}_A$  and ($\rm{Sn}_A$+ $\rm{Sn}_B$) and their corresponding photoemission cross-sections. 
~Note that following similar procedure for 0.9 ML Sn deposited on Au(111) at RT, we find the composition to be AuSn, in agreement with literature\cite{Barthes81}. Thus, clearly, besides the surface structure, the composition of the surface alloy also changes with $T_S$.

 It is important to note that the occurrence of the Dirac-like linear bands in Fig.~\ref{arpes} is specific to the (2113) phase with composition of Au$_2$Sn. The ARPES for the other phases with different surface structure and composition studied by us do not show the linear bands (Fig.~S3)\cite{supplement}. %  Neither at lower coverages at  T$_S$ of 423 K or at $\approx$ 1 ML coverage at other T$_S$, this (2113) phase does not form and the linear bands are not observed (Fig. S2).  
       ~       DFT has been extensively used to understand the electronic structure of the 2D materials. %calculate the band structure of the (2113) phase to understand the origin of the linear bands. 
~	However,  in order to perform DFT, a starting structural model for the (2113) phase is required.  % that can occur over multiple layers\cite{Barthes81,Zhang91}. Moreover, 
       ~Although we have determined the  unit cell,  the task of obtaining the atomic positions is complicated due to surface alloying and is outside the scope of the present work. If bare Au(111) surface is considered, the unit cell would comprise of 5 Au atoms in the regular fcc positions  with the lattice parameters {\bf $b_1$}= 4.89\AA, {\bf $b_2$}= 7.61\AA~ and  $\gamma$= 70.9$^{\circ}$. But, for Au$_2$Sn, a 2:1 ratio of the Au and Sn atoms is not satisfied with 5 atoms in the unit cell. % if we consider substitution of Au by Sn. %A possibility is that for 1 ML Sn, this would mean presence of 3 layers of Au$_2$Sn  (1/3 Sn ML for each layer), and thus 15 atoms of the (2113) unit cell can satisfy the 2:1 ratio with 10 Au atoms and 5 Sn atoms. However, considering $^{15}$C$_5$, there are 42042 possible ways of substituting 5 Sn atoms in 15 Au sites. A way out is of this quagmire is to consider an 
       ~So, we have considered an extra Sn atom noting that %to satisfy this requirement.  %the that should be accommodated to commensurate with the requisite ratio of 2:1. 
%       ~It is worth  mentioning here that since 
~the size of the unit cell ({\bf $b_2$}= 8.4$\pm$0.3\,\AA~ obtained from STM) is larger than the unrelaxed bulk terminated unit cell with {\bf $b_2$}= 7.61\AA. %, we pursue our theoretical calculations with $b_2$= 8.2\,\AA and $b_1$= 5.37\,\AA. 
       ~Among a few structures we have probed, we present here an  atomic arrangement following the  Lieb lattice that  has been shown to host a topologically non-trivial phase with a single Dirac point per unit cell along with a dispersionless band through it\cite{Weeks10}.  However, in order to be consistent with experiment, we take a non-primitive  Lieb lattice with 6 atoms per unit cell and modified to be oblique with $\gamma$= 70.9$^{\circ}$ rather than being rectangular\cite{Weeks10}. To retain the inversion symmetry, we assume two Sn atoms to be at the center and corner positions, while the four Au atoms occupy intermediate positions forming a parallelogram with subtended  angle of $\gamma$ (Fig.~S4(a))\cite{supplement}.      The calculated band structures reveal that only for this model, %along $\overline{\Gamma}$$\overline{M^{\prime}}$ in  
~        a pair of linear bands   meet at %that are symmetric with respect to 
       ~the $\overline{\Gamma}$ point, but at about 2 eV above $E_F$ and shows a small gap of 130 meV (Fig.~S4(b,c)\cite{supplement}). These bands originate primarily from the Sn $p$ states with some admixture of Au $s$ states. %These bands originate primarily from the Sn $p$ states with some admixture of Au $s$ states.  %The Dirac point occurs above $E_F$ at about 2~eV  %whereas in experiment we observe it at $E_F$.  
       ~However, our experimental data show that the linear bands meet  at $E_F$  and also the two shallow parabolic bands crossing the linear bands are not observed (Fig.~\ref{arpes}).       
The reasons for such qualitative disagreement with ARPES could be related to the fact that the (2113) phase occurs only at high temperature and thus it could be a metastable phase, while DFT calculates the lowest energy  ground state at zero temperature. Most importantly, we have used only a notional model structure consistent with the experimentally determined unit cell. The disagreement may  also be attributed to that. %Indeed, we find from our calculation that if $\gamma$ is relaxed, a total energy  minimum  is obtained for $\gamma$= 90$^{\circ}$. %, but in this case the linear bands are not observed.
~We believe that the complete   structure of the (2113) phase needs to be determined experimentally and used as input to the DFT calculation for realistic determination of its electronic band structure. %However, our preliminary DFT calculations show the importance of oblique symmetry of the Lieb lattice in obtaining Dirac-like linear bands in this two component system.} % should be performed for high temperature where this phase is stable.      

%\section{Conclusions}
To conclude, we have identified  Dirac-like linear bands at the zone center in Au$_2$Sn surface alloy that has an oblique unit cell  described by $M = \left( \begin{smallmatrix} 2&1\\ 1&3 \end{smallmatrix} \right)$. This is thus an example of  Dirac cone in a non-honeycomb surface alloy. The linear bands forming the Dirac cone have Fermi velocity comparable to that of graphene.  %The composition of the surface alloy is determined from core level photoelectron spectroscopy. 
%~Our DFT calculation shows presence of  linear bands for this binary alloy represented by a model structure, namely modified Lieb lattice with oblique symmetry. 
~ It is surprising  that exotic electronic structure is possible even in %unexpected 
complicated surface alloys. The present results  will rejuvenate the search for 2D quantum materials that are  important for high speed electronic devices.

D.P. and A.C. thanks P.A. Naik, A. Banerjee  for support and encouragement and the Computer Centre of RRCAT, Indore  for providing the computational facility.

%\newpage
\setcounter{figure}{0}
\renewcommand{\figurename}{Fig.~S}

\begin{center}
	{\bf Supplementary material for the paper entitled "Dirac cone in a non-honeycomb surface alloy" }
~~\\
~~\\
Pampa Sadhukhan$^1$, Dhanshree Pandey$^{2,3}$, Vipin  Kumar Singh$^1$, Shuvam Sarkar$^1$, Abhishek Rai$^1$, Kuntala Bhattacharya$^4$,  Aparna Chakrabarti$^{2,3}$ and Sudipta Roy Barman$^{1}$
~~\\
~~\\
$^1$UGC-DAE Consortium for Scientific Research, Khandwa Road, Indore 452001, Madhya Pradesh, India\\
$^2$Homi Bhabha National Institute, Training School Complex, Anushakti Nagar, Mumbai  400094, Maharashtra, India\\
$^3$Theory and Simulations Laboratory,  Raja Ramanna Centre for Advanced Technology, Indore 452013, Madhya Pradesh, India\\
$^4$Department of Physics, Indian Institute of Space Science and Technology, Thiruvananthapuram 695547, Kerala, India\\
~~\\
~~\\
 This Supplementary material contains four figures.
\end{center}

\begin{figure*}[b]
	%\centering
	%\epsfxsize=120mm\\
	\begin{center}
		\includegraphics[width=140mm,keepaspectratio]{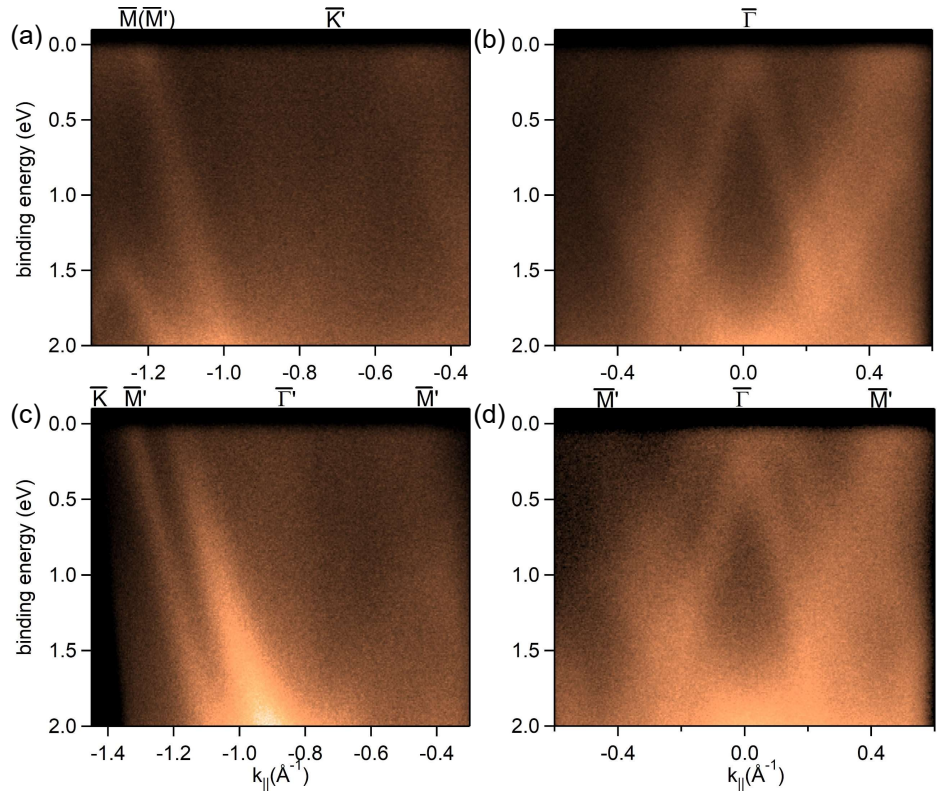}%% Source: anlysis_final/Layout30,  
		\caption{ARPES spectra of the (2113) phase along $\overline\Gamma$$\overline{M}$ direction (a) over an extended range up to $\overline{M}$ point and (b) around $\overline\Gamma$ point.  ARPES spectra along $\overline\Gamma$$\overline{K}$ direction (c) over an extended range up to $\overline{K}$ point and (d) around $\overline\Gamma$ point.}
		%\label{S1}
	\end{center}
\end{figure*}

\begin{figure*}[h]
	%\centering
	%\epsfxsize=120mm\\
	\begin{center}
		\includegraphics[width=160mm,keepaspectratio]{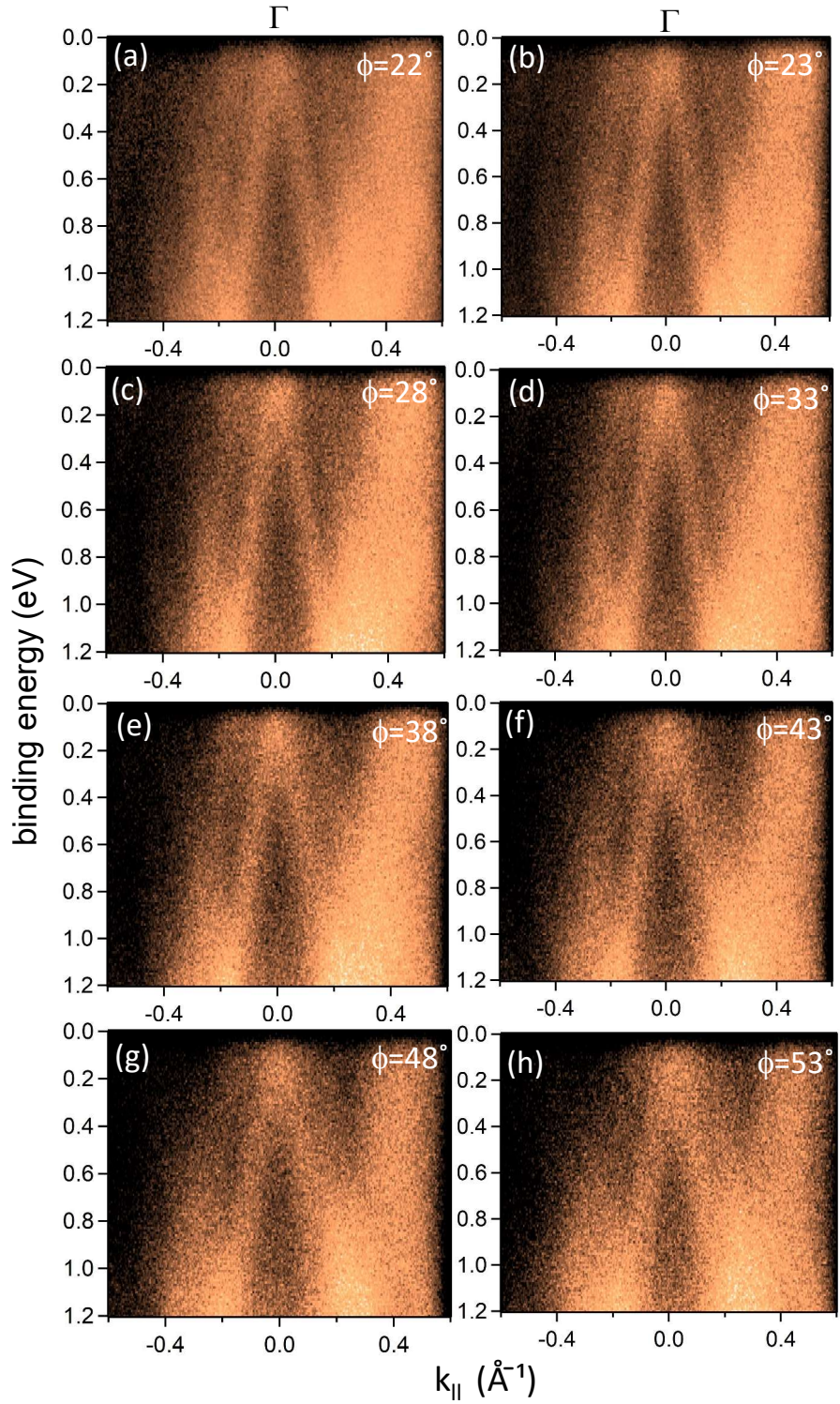}%% Source: anlysis_final/Layout30,  
		\caption{ARPES spectra around $\overline\Gamma$ point from $\overline\Gamma$$\overline{M}$ to $\overline\Gamma$$\overline{K}$ direction at azimuthal angle $\phi$= (a) 22$^{\circ}$, (b) 23$^{\circ}$ ($\overline\Gamma$-$\overline{M}$), (c) 28$^{\circ}$, (d) 33$^{\circ}$, (e) 38$^{\circ}$, (f) 43$^{\circ}$, (g) 48$^{\circ}$ and (h) 53$^{\circ}$ ($\overline\Gamma$-$\overline{K}$).}
		%\label{S2}
	\end{center}
\end{figure*}

\begin{figure*}[h]
	%\centering
	%\epsfxsize=120mm\\
	\begin{center}
		\includegraphics[width=155mm,keepaspectratio]{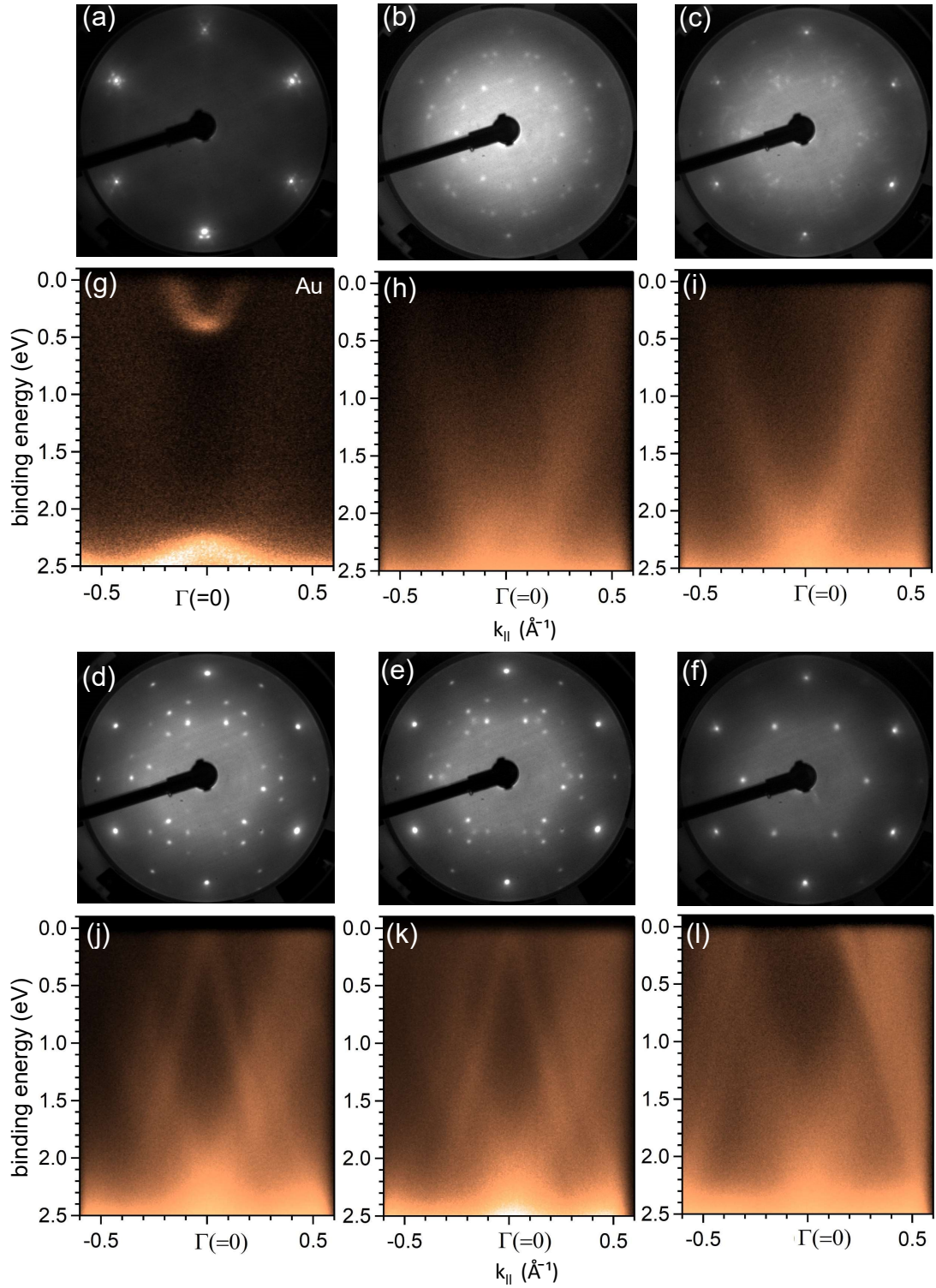}%% Source: anlysis_final/Layout30,  
		\caption{Low energy electron diffraction pattern and ARPES spectrum of (a, g)  Au(111) and  0.9 ML Sn/Au(111) deposited at (b, h) $T_S$= 300~K forming a $p$(3$\times$3)-$R$15$^{\circ}$ pattern,  (c, i) $T_S$= 383~K forming a mixed phase (d, j) $T_S$= 413~K forming the (2113) phase, (e, k) $T_S$= 443~K forming a  $\sqrt{3}$$\times$$\sqrt{3}$ phase mixed with the (2113) phase  and (f, l) $T_S$= 493~K forming $\sqrt{3}$$\times$$\sqrt{3}$ phase, respectively.}
	\end{center}
\end{figure*}

\begin{figure*}[h]
	\centering
	\vskip -20mm
	\includegraphics[scale=0.6]{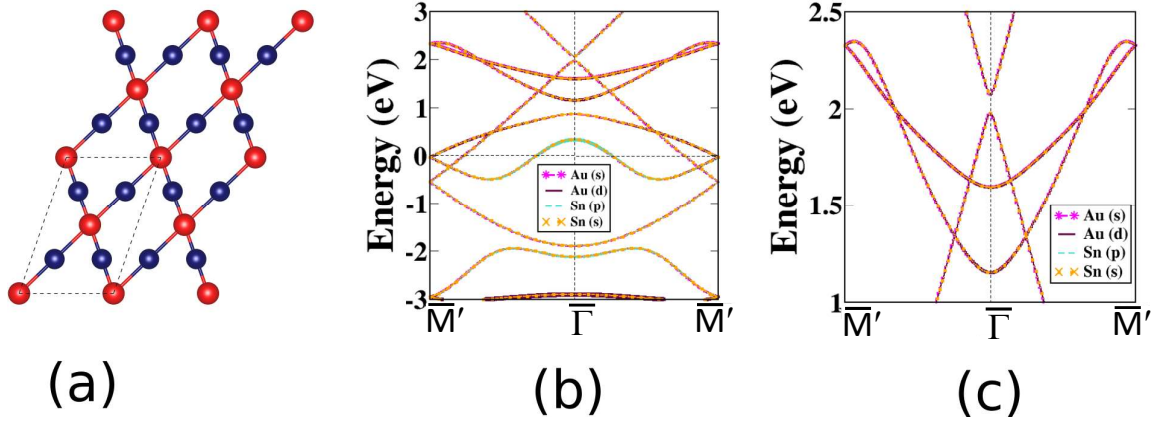} 
	\vskip -40mm
	\caption{(a) The structure of the  oblique Lieb lattice after atom position relaxation, the unit cell is marked by black dashed line, Au and Sn atoms are shown by dark blue and red filled circles, (b) The corresponding band structure  is shown along $\overline{\Gamma}$-$\overline{M^{\prime}}$ direction, the contribution of the Au and Sn states to each band is indicated, (c) the linear bands are shown in an expanded scale.}
	\label{S4}	
\end{figure*}

\end{document}